\documentclass[aps,prl,amsmath,twocolumn,superscriptaddress,letterpaper,floatfix]{revtex4-1}
\usepackage{graphicx,color}
\usepackage{verbatim}
\usepackage{amssymb}   
\usepackage{amsmath}
\usepackage{amsfonts}
\usepackage{mathdots}
\usepackage{braket}
\usepackage{bm}
\usepackage{float}
\usepackage{bbm}

\usepackage{soul}
\usepackage{hyperref}

\begin{document}
\title{Topological altermagnetic Josephson junctions}
\author{Grant Z. X. Yang}
\affiliation{Department of Physics, Hong Kong University of Science and Technology, Clear Water Bay, Hong Kong, China}

\author{Zi-Ting Sun}\thanks{zsunaw@connect.ust.hk} 
  \affiliation{Department of Physics, Hong Kong University of Science and Technology, Clear Water Bay, Hong Kong, China} 	

\author{Ying-Ming Xie}
 \affiliation{RIKEN Center for Emergent Matter Science (CEMS), Wako, Saitama 351-0198, Japan}

  \author{K. T. Law} \thanks{phlaw@ust.hk}

  \affiliation{Department of Physics, Hong Kong University of Science and Technology, Clear Water Bay, Hong Kong, China}

\date{\today}

\bigskip

\begin{abstract}
Planar Josephson junctions are pivotal for engineering topological superconductivity, yet are severely hindered by orbital effects induced by in-plane magnetic fields. In this work, we introduce the generic topological altermagnetic Josephson junctions (TAJJs) by leveraging the intrinsic spin-polarized band splitting and zero net magnetization attributes of altermagnets. Our proposed TAJJs effectively mitigate the detrimental orbital effects while robustly hosting Majorana zero modes (MZMs) at both ends of the junction. Specifically, we demonstrate that MZMs emerge in $d_{x^2-y^2}$-wave TAJJs but vanish in the $d_{xy}$-wave configuration, thereby establishing the orientation angle $\theta$ of altermagnet as an emergent control parameter of topological superconductivity. The distinct spin-polarization of the MZMs provides an experimental signature for the spin-resolved measurement. Furthermore, by harnessing the synergy between the $d_{x^2-y^2}$-wave altermagnet and anisotropic superconductivity, our proposal extends to high-$T_c$ SC platforms naturally. Overall, this work establishes altermagnets as a versatile paradigm for realizing topological superconductivity, bridging conceptual innovations with scalable quantum architectures devoid of orbital effects and stray fields.

\end{abstract}
\bigskip

\maketitle

\label{sec:intro}

\emph{Introduction.}\textemdash
The interplay between superconductivity (SC) and magnetism—particularly as the segment for the lifted Kramers spin degeneracy (LKSD) via the proximity effect—has provided a rich landscape for unveiling Majorana zero modes (MZMs), the prime candidates for realizing non-Abelian statistics in topological quantum computation \cite{fu_TI_MEM,SP-STM_1,SP-STM_2,SP-STM_3,magnet_MEM_6,magnet_MEM_5,magnet_MEM_4,magnet_MEM_1,Linder_topojj_ferro, Kitaev_1D_p_wave2000,Beenakker_MEM_review_2011,Alicea2012, KITAEV20032,nat_rev_mat_1,nat_rev_mat_2}. 
Planar Josephson junctions, fundamental components in superconducting devices \cite{Kockum2019_JJ_book}, have been proposed as promising platforms for hosting MZMs at their ends. This is achieved by tuning the phase difference $\phi$ between the superconducting leads to $\pi$, while an in-plane magnetic field combined with proximitized Rashba spin-orbit coupling (SOC) induces a topological gap in the Andreev bound state (ABS) spectrum \cite{topoJJ,topoJJ_exp_1,topoJJ_exp_2, topoJJ_2, XieYM_PRL,2D_MBS_platform,topoJJ_3, topoJJ_4, sun2024anomalous}.
However, the reliance on either magnetic fields or ferromagnets introduces significant challenges. Both the strong stray fields in ferromagnets and orbital effects from the in-plane magnetic field will inevitably suppress the SC, thereby undermining the topological gap \cite{Proximity_effects_ferro,topoJJ,topoJJ_exp_2}.

\begin{figure}[h]
    \centering
    \includegraphics[scale=0.18]{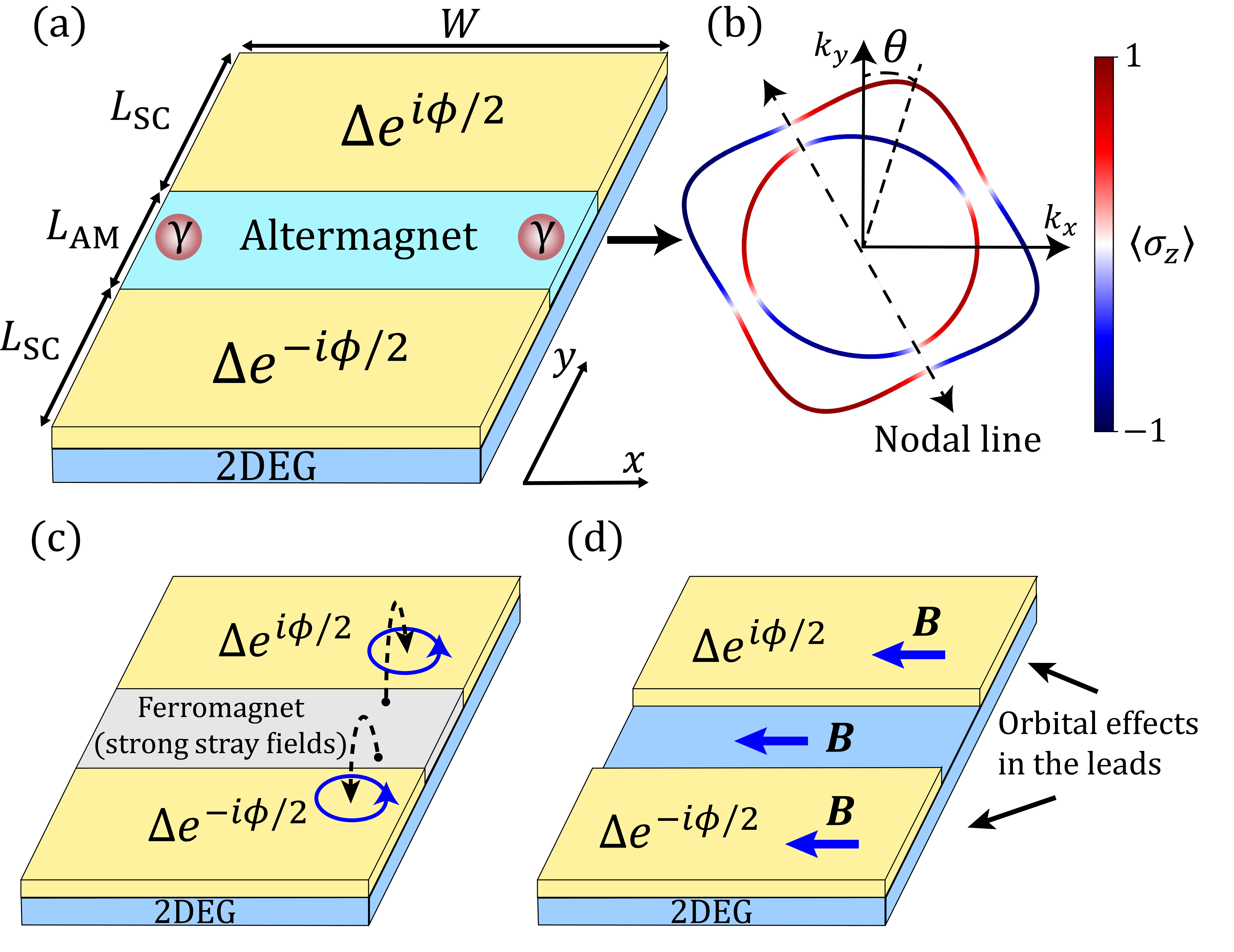}
    \caption[short]{(a) Schematic of a TAJJ atop a two-dimensional electron gas (2DEG) hosting the MZMs labelled by $\gamma$. (b) Fermi surface of an altermagnet in the presence of Rashba SOC, where the color bar indicates the normalized out-of-plane spin polarization. Rashba SOC opens gaps along the nodal lines, where the out-of-plane spin polarization vanishes. (c) A ferromagnetic Josephson junction showing strong stray fields. (d) A planar Josephson junction under an in-plane magnetic field, where orbital effects in the leads significantly affect the superconducting gap.}
    \label{fig:1}
\end{figure}

Recent advances in magnetism have led to the discovery of an unconventional phase known as altermagnetism (AM), characterized by anisotropic spin-polarization yet a zero net magnetization~\cite{altermagnet_prx_1,altermagnet_prx_2,song2025altermagnets}. These unique attributes enable the design of stray-field-free Josephson junctions, where a strong non-relativistic LKSD can be achieved at zero external field \cite{anti_stray_field_free,stray_fields,stray_field_free, ma2021multifunctional}. The potential for topological SC in heterostructures and metallic systems with AM has been discussed in earlier theoretical studies \cite{Jennifer_MEM_AM2024, mondal2025distinguishing,yanzhongbo_topo_sc_metal,li2023majorana,li2024creation,sun2025pseudo}. Moreover, the integration of $d$-wave altermagnetic materials into Josephson junctions has received attention. These works have reported anomalous current-phase relations, spin-splitter and spin-filter effects, and phase-shifted ABS spectra exhibiting gap opening at $\phi=\pi$ \cite{dcamjj,Beenakker_alterjj,amjj_1,amjj_3,amjj_4,sun2024tunable, zhang2024finite, hu2025nonlinear, fukaya2025josephson}. Notably, spin-resolved photoemission spectroscopy demonstrates the non-relativistic LKSD energy scale can reach half an eV in altermagnets like MnTe and CrSb \cite{LKSDexp,LKSDexp_2,LKSDexp_3,LKSDexp_4}, while tunable altermagnetic order via crystal symmetry engineering has been achieved in CrSb \cite{zhou_manipulation_2025}. Moreover, a $d$-wave metallic altermagnet, $\text{KV}_2\text{Se}_2\text{O}$, has been revealed experimentally at room-temperature \cite{jiang2025metallic}. These developments pave the way for realizing unconventional altermagnet-based Josephson junctions that features large LKSD with vanishing net magnetization.

In this work, we propose a zero-field platform for achieving topological SC based on altermagnetic planar Josephson junctions, namely topological altermagnetic Josephson junctions (TAJJs). The strong proximity effect between a superconductor and a two-dimensional electron gas (2DEG), supported by mature experimental techniques \cite{soc_proximity1,soc_proximity2,soc_proximity3}, underpins the minimal design of TAJJs [Fig.~\ref{fig:1}(a)]. The device comprises two $s$-wave superconductor leads with a finite phase difference $\phi$ and a $d$-wave altermagnet placed atop a 2DEG with strong Rashba SOC. Breaking spin-degeneracy is indispensable for emulating the physics of spinless $p$-wave Kitaev chains and realizing non-Abelian statistics~\cite{Kitaev_1D_p_wave2000,Sarma_non_abelian}, although time-reversal symmetry (TRS) breaking typically harms SC \cite{BCS_theory}. Therefore, it is crucial that altermagnet can be fabricated solely within the junction, and MZMs are regarded as the special ABSs at the ends \cite{MZM_as_ABS}. This approach avoids the pitfalls of applying an in-plane magnetic field, which cannot be confined [Fig.~\ref{fig:1}(d)] and the orbital effects dampen or even close the superconducting gap \cite{topoJJ}. Besides, the zero net magnetization renders TAJJs stray-field-free in contrast to ferromagnetic Josephson junctions in Fig.~\ref{fig:1}(c) \cite{stray_fields}. Additionally, the orientation angle $\theta$ of altermagnet (see Fig.~\ref{fig:1}(b)) serves as a parameter controlling the emergence of topological SC, can be {changed} to distinguish the MZMs from accidental zero-energy end states experimentally \cite{Vic_liujie_disorder}.

In the following sections, we present a comprehensive theoretical analysis of topological SC in $d$-wave TAJJs. By examining the spin-resolved ABS spectra and constructing topological phase diagrams—via calculating $\mathbbm{Z}_2$ ($\mathbbm{Z}$) invariants in the D (BDI) symmetry classes for one-dimensional systems—we elucidate the conditions conducive to the emergence of MZMs. Moreover, the proposed platform is extendable to high-$T_c$ superconductors, where the interplay between $d_{x^2-y^2}$-wave AM and anisotropic SC can drive the system into a topological superconducting phase robustly with high critical temperatures.

\begin{figure}[t]
    \hspace*{-0.2 cm}
    \includegraphics[scale=0.33]{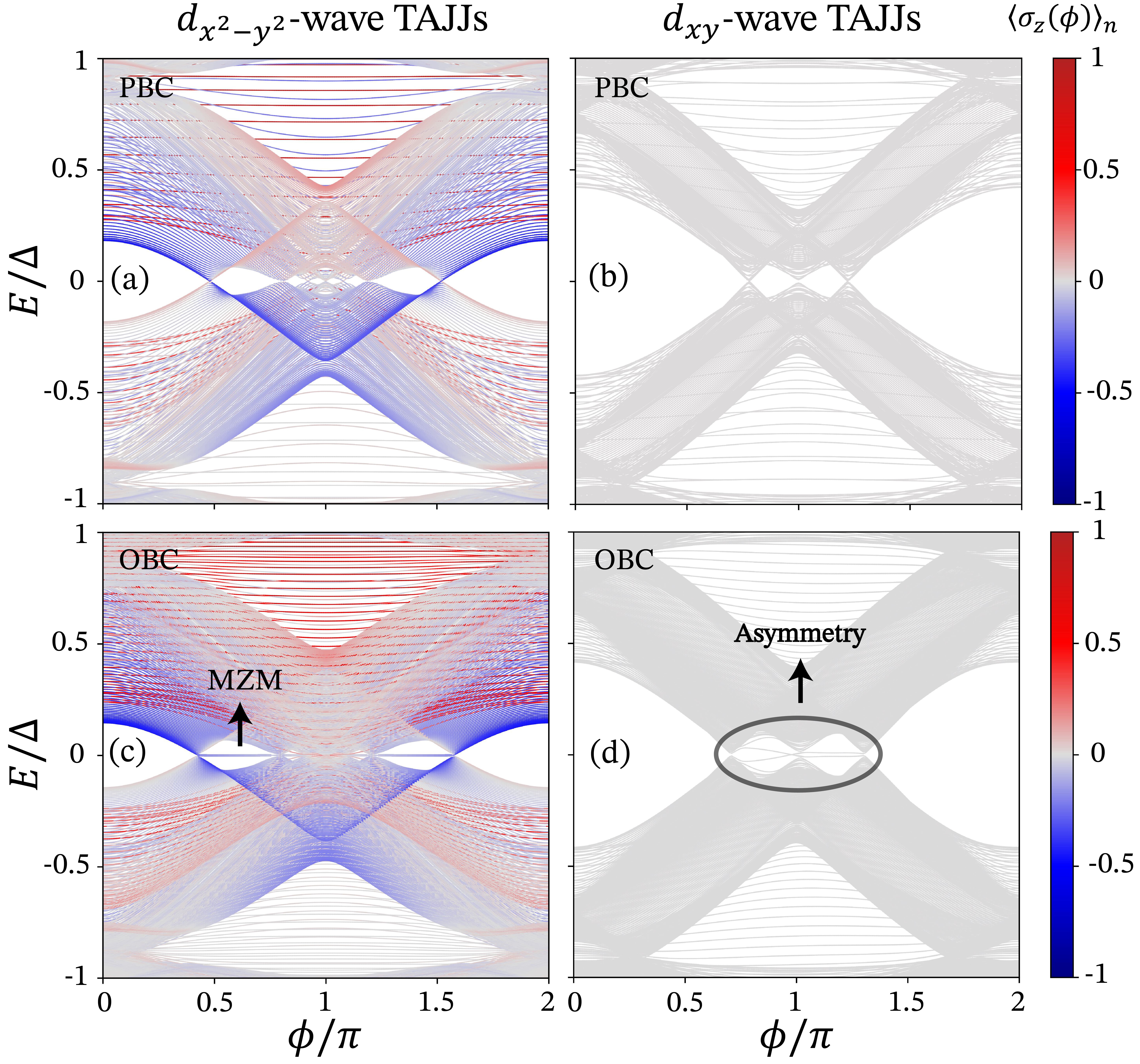}
    \caption{ABS spectra of TAJJs calculated in (a,b) periodic and (c,d) open boundary conditions. 
    The color indicates the out-of-plane spin-polarization profile $\langle\sigma_z(\phi)\rangle_n$ for each ABS $\ket{\psi_n(\phi)}$, 
    normalized to $\pm1$. In $d_{x^2-y^2}$-wave TAJJs, MZMs emerge and ABSs exhibit strong spin-polarization. 
    In contrast, for $d_{xy}$-wave TAJJs, both MZMs and spin-polarization are absent due to the degeneracy of states $\ket{\psi_n(\phi,\pm k_x)}$ with opposite out-of-plane spin-polarization.} 
    \label{fig:2}
\end{figure}

\emph{Continuum model.}\textemdash
We investigate TAJJs consisting of two $s$-wave superconducting leads separated by a $d$-wave altermagnetic weak link, as illustrated in Fig.~\ref{fig:1}(a). The superconducting leads maintain a finite phase difference $\phi$, and Rashba SOC is proximitized across the entire system. We consider a geometry where both the junction width $W$ and superconducting lead length $L_{\text{SC}}$ approach infinity, while the altermagnetic region length $L_{\text{AM}}$ remains {finite}. The Bogoliubov-de Gennes (BdG) Hamiltonian describing this system in the Nambu spinor basis $\Psi = (\psi_\uparrow, \psi_\downarrow, \psi^\dagger_\downarrow, -\psi^\dagger_\uparrow)^T$ is given by:
\begin{equation}
    \begin{split}
    &H_{\text{BdG}} = \left[t(\hat{k}_y^2+k_x^2)-4t-\mu+2\alpha(k_x\sigma_y-\hat{k}_y\sigma_x)\right]\tau_z\\
    &+ M(\theta)\sigma_z\Theta(L_{\text{AM}}/2-|y|) + \Delta(y)\tau_+ + \Delta^*(y)\tau_-, \\
    \end{split}
    \label{BdG_H}
\end{equation}
where $\sigma$ and $\tau$ are Pauli matrices acting on spin and particle-hole space respectively, with $\tau_{\pm}=(\tau_x\pm i\tau_y)/2$. The kinetic energy term is parameterized by $t$ (with $\hat{k}_y=-i\partial_y$ representing the momentum operator), $\alpha$ denotes the Rashba SOC strength, $\mu$ is the chemical potential, and $\Theta$ is the Heaviside step function that restricts AM and SC terms to their respective regions.
The superconducting $s$-wave pairing potential in the leads is defined as:
\begin{equation}
\Delta(y) = \Delta e^{i\text{sgn}(y)\phi/2}\Theta(|y|-L_{\text{AM}}/2),
\label{pairing}
\end{equation}
where $\Delta$ is the pairing amplitude. 
The $d$-wave altermagnetic order parameter is expressed as:
\begin{equation}
M(\theta) = m\left[(k_x^2-\hat{k}_y^2)\cos{2\theta}+4k_x\hat{k}_y\sin{2\theta}\right], 
\end{equation}
where $\theta \in [0,\pi/4]$ parameterizes the orientation of the $d$-wave AM. The cases $\theta=0$ and $\theta=\pi/4$ correspond to $d_{x^2-y^2}$ and $d_{xy}$-wave AM, respectively, as shown in Fig.~\ref{fig:1}(b). Within the altermagnetic region, the AM distorts the spin-degenerate Fermi surfaces into two orthogonal ellipses with opposite out-of-plane spin-polarization~\cite{altermagnet_prx_1,altermagnet_prx_2}. Importantly, along the nodal lines, this spin-polarization vanishes and TRS is preserved, while Rashba SOC further induces gaps at the nodes of the two $d$-wave altermagnetic Fermi surfaces.

\begin{figure*}[tbp]
    \centering
    \hspace*{-0.3cm}
    \includegraphics[scale=0.34]{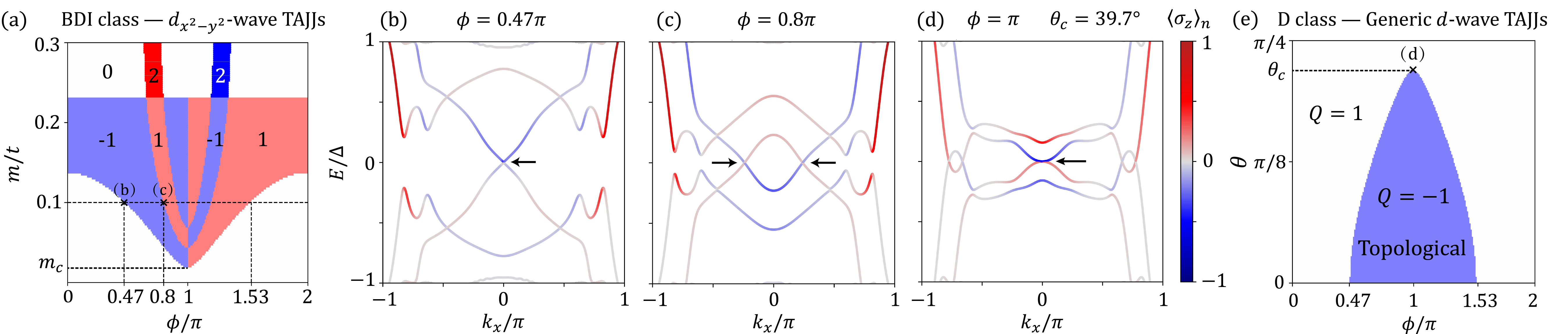}
    \caption{Topological characterization of TAJJs: (a) Phase diagram of $d_{x^2-y^2}$-wave TAJJs in BDI symmetry class as a function of parameters $(\phi, m)$ and (e) generic $d$-wave TAJJs in D symmetry class as a function of $(\phi, \theta)$. 
    Regions with odd (even) winding number $w$ correspond to topological (trivial) phases.
    (b-d) ABS spectra for fixed $\phi$ values in $k_x$-space, revealing band inversions at the boundaries between topological sub-regions.
    The dashed lines in (a) and (e) are obtained from Eq. (\ref{abs_scattering}) without normal reflection.
    The AM strength in (e) is $m_0/t=0.1$, and the critical points in (a) and (e) satisfy: $m_c\approx m_0\cos{2\theta_c}$.}
    \label{fig:3}
\end{figure*}

\emph{{Spin-{resolved} ABS spectra.}}\textemdash
For numerical simulations, we discretize the continuum model (Eq.~\ref{BdG_H}) into a tight-binding model (see Supplemental Material (SM) I for details \cite{Supplemental}). To obtain both the energy spectra and wavefunctions of ABS, we diagonalize the BdG Hamiltonian under both periodic boundary conditions (PBC) and open boundary conditions (OBC). Our calculations employ the following parameter values: $t=1$, $\Delta=0.3$, $\alpha=0.25$, $m=0.1$, $\mu=-0.5$, $L_{\text{AM}}=3$, $L_{\text{SC}}=15$, and $W=200$. To investigate the out-of-plane spin-polarization characteristics of ABSs induced by AM, we define the corresponding spin-polarization profile for each ABS $\ket{\psi_n(\phi)}$ as $\langle\sigma_z(\phi)\rangle_n=\bra{\psi_n(\phi)}\sum_i\sigma^i_z\ket{\psi_n(\phi)}$, where $\sum_i$ denotes summation over all lattice sites. This spin-polarization is represented by the color scale in Fig.~\ref{fig:2}(a-d), with gray color indicating non-spin-polarized ABSs.

Our analysis reveals two noteworthy features in the ABS spectra:
(1) The emergence of spin-polarized MZMs in $d_{x^2-y^2}$-wave TAJJs [Fig.~\ref{fig:2}(c)]. (2) The absence of both MZMs and out-of-plane spin-polarization in the $d_{xy}$-wave TAJJs [Fig.~\ref{fig:2}(b) and (d)]. We next explain the underlying reasons for the above features. 
The emergence of topology in the TAJJs is determined only by the band inversions at $k_x=0$ \cite{topoJJ,Kitaev_1D_p_wave2000}, 
where the ABS spectrum of $d_{x^2-y^2}$-wave TAJJs without normal reflection is given below by scattering matrix formalism [see SM III \cite{Supplemental}]
\begin{equation}
    E(\phi)=\pm\nu \Delta \cos\left(\frac{\phi}{2}\mp\nu\frac{\pi}{2}\frac{E_{\text{AM}}}{E_{\text{T}}}\right)
    \label{abs_scattering}
\end{equation}
Here $\nu=\pm1$ indicates the out-of-plane spin-polarization direction. The phase shift $E_{\text{AM}}=m k_F^2$ is given in the units of Thouless energy $E_{\text{T}}=(\pi/2)v_F/L_{\text{AM}}$ and $k_F=\sqrt{(\mu+4t)/t}$. The $d_{x^2-y^2}$-wave altermagnetic LKSD opens a gap in ABS spectrum at $k_x=0$ and Rashba SOC furtherly protects the gap for the modes with $k_x\ne 0$ \cite{Supplemental,topoJJ}, rendering system to be topological for $(\phi-\pi) \in  (E_{\text{AM}}/E_{\text{T}})\times[-\pi, \pi]$.
Meanwhile, the order parameter of $d_{xy}$-wave AM vanishes if $k_x=0$: $M(\theta=\pi/4) \propto 2\sin{k_x} (-i\partial_y)|_{k_x=0,\pi}=0$. Consequently, topological SC in TAJJs arises solely from the $d_{x^2-y^2}$-wave AM. As shown in Fig.~\ref{fig:2}(c), the resulting MZMs exhibit distinct out-of-plane spin-polarization, which can be detected through selective equal-spin Andreev reflection measurements \cite{SESAR}.
In the case of $\theta=\pi/4$ ($d_{xy}$-wave AM), where the supercurrent flows along the nodal direction, a fundamental degeneracy emerges between ABSs $\ket{\psi_n(k_x,\uparrow)}$ and $\ket{\psi_n(-k_x, \downarrow)}$ for a fixed $\phi$, since $M(\theta=\pi/4)\sigma_z \propto k_x\hat{k}_y\sigma_z$. As a result, the net spin polarization $\langle\sigma_z(\phi)\rangle_n$ for each Andreev level vanishes in $d_{xy}$-wave TAJJs, as the spin-polarization of degenerate states exactly cancel each other. This bound-state degeneracy is a general property of $d_{xy}$-wave AM and can be demonstrated analytically in altermagnetic ribbons under OBC \cite{dxy_degeneracy}. 
Aside from these, we also analyze the local magnetization profiles $\langle\sigma_z^i(\phi)\rangle = \sum_{E_n<0}\bra{\psi_n(\phi)}\sigma_z^i\ket{\psi_n(\phi)}$ defined on each lattice site $i$ in SM VIII \cite{Supplemental}. These profiles reveal vanishing out-of-plane bulk magnetization but various edge magnetization patterns in TAJJs depending on $d$-wave AM orientations, akin to the spin-splitter effect discussed in quasiclassical theory \cite{splitter_effect}.

\emph{Symmetry and topological classification.}\textemdash
The symmetry properties of TAJJs determine their topological classification and associated invariants. For generic $d$-wave orientations ($0 < \theta \leq \pi/4$), TAJJs belong to symmetry class D with particle-hole symmetry $P = \sigma_y\tau_y\mathcal{K}$ (where $\mathcal{K}$ denotes complex conjugation operator) and are characterized by a $\mathbb{Z}_2$ topological invariant. In these systems, both the conventional time-reversal symmetry $\mathcal{T} = i\sigma_y\mathcal{K}$ and mirror symmetry in the $x$-$z$ plane $M_y = (y \rightarrow -y) \times i\sigma_y$ are broken by $d_{x^2-y^2}$-wave AM and the finite phase difference between the leads.
However, for the special case of $d_{x^2-y^2}$-wave TAJJs ($\theta = 0$), an effective time-reversal symmetry $\tilde{\mathcal{T}} = M_y\mathcal{T}$ with $\tilde{\mathcal{T}}^2 = 1$ is preserved. This additional symmetry elevates the system to the BDI symmetry class, characterized by an enriched $\mathbb{Z}$ topological invariant. In this case, the system also possesses chiral symmetry $C = M_y\tau_y$ \cite{topoJJ}. When any $d_{xy}$-wave component is present ($\theta \neq 0$), both effective and conventional time-reversal symmetries are broken, returning the system to class D.

In addition, an interesting asymmetry in the ABS spectrum appears specifically for $d_{xy}$-wave TAJJs under OBC [Fig.~\ref{fig:1}(d)] but is absent under PBC, resulting in an edge Josephson diode effect \cite{JDE_1, JDE_2, JDE_3}. For bulk transport, the absence of nonreciprocal behavior in $d$-wave TAJJs (regardless of $\theta$) can be understood from the symmetry constraint $C_{2z}H_{\text{BdG}}(k_x,\phi)C_{2z}^{-1} = H_{\text{BdG}}(k_x,-\phi)$ under PBC, where $C_{2z} = (x,y \rightarrow -x,-y) \times \sigma_z$. This holds even though both TRS and inversion symmetry are broken \cite{AMSDE}. Under OBC, however, the $C_{2z}$ constraint is violated by the interplay between local edge magnetization of $d_{xy}$-wave AM and Rashba SOC, enabling the edge Josephson diode effect \cite{zhang2025observation} near the boundaries (see SM IX \cite{Supplemental}).

\emph{Topological phase diagram of TAJJs.}\textemdash 
As established previously, $d_{x^2-y^2}$-wave TAJJs ($\theta = 0$) belong to the BDI symmetry class with enhanced symmetry protection. To determine the phase diagram, we employ PBC along the $x$-direction (making $k_x$ a good quantum number) and compute the winding number $w$ as a function of parameters $(\phi, m)$ following the methodology of Refs.~\cite{winding_calculation_prl,topoJJ, Supplemental}. Odd (even) values of $w$ correspond to topological (trivial) phases. The resulting phase diagram in Fig.~\ref{fig:3}(a) exhibits a diamond-like shape similar to systems with in-plane magnetic fields \cite{topoJJ}, the dashed line is calculated based on Eq. (\ref{abs_scattering}) without normal reflection. As the AM strength $m$ increases, the topological region expands in $\phi$-space until MZMs emerge without requiring fine-tuning of $\phi$, and the system remains topological for the entire range $\phi \in [0,2\pi]$. Distinctively, the phase diagram contains two sub-regions with opposite winding numbers $w = \pm 1$ and is symmetric with respect to $\phi = \pi$. To further investigate these sub-regions, we calculate the ABS spectra as functions of $k_x$ at fixed $\phi$ values, shown in Fig.~\ref{fig:3}(b-d). At the phase transition point $(\phi = 0.47\pi, m/t = 0.1)$ in Fig.~\ref{fig:3}(b), a single gap closes at $k_x = 0$, driving the system into the topological phase. When $\phi$ reaches approximately $0.8\pi$, the winding number jumps from $w = -1$ to $w = 1$, corresponding to the simultaneous closing of two gaps at two {opposite} finite momenta ($k_x \neq 0$) in Fig.~\ref{fig:3}(c). The system remains topological but with an opposite odd winding number, and each intersection between distinct sub-regions corresponds to a gap-closing point in the ABS spectra as a function of $\phi$ [Fig.~\ref{fig:2}(a,c)].

For generic $d$-wave TAJJs ($\theta \neq 0$) in symmetry class D, the particle-hole symmetry $P$ guarantees a well-defined $\mathbb{Z}_2$ topological invariant as
$Q = \text{sgn}\left[\text{Pf}\{H(k_x=\pi)\sigma_y\tau_y\}/{\text{Pf}\{H(k_x=0)\sigma_y\tau_y\}}\right]$,
and $\text{Pf}\{\cdots\}$ denotes the Pfaffian and the topological phases reside in the region with $Q = -1$ \cite{winding_calculation_prl,winding_calculation_2,XieYM_winding_number}. The resulting phase diagram as a function of $(\phi, \theta)$ is presented in Fig.~\ref{fig:3}(e).
For one-dimensional systems in the class BDI, the $\mathbb{Z}_2$ invariant corresponds to the parity of the $\mathbb{Z}$ invariant \cite{winding_calculation_prl}. Consequently, the topological regions for $d_{x^2-y^2}$-wave TAJJs ($\theta = 0$) in Fig.~\ref{fig:3}(a) and Fig.~\ref{fig:3}(e) match exactly for the same parameters. As $\theta$ increases from $0$ toward $\pi/4$, the AM in TAJJs rotates from $d_{x^2-y^2}$-wave to $d_{xy}$-wave, $\theta$ effectively controls the magnitude of the topologically relevant $d_{x^2-y^2}$-wave AM as $m(k_x^2-k_y^2)\cos(2\theta)$. This transition drives the system from topological to trivial phases for $\phi$ near $\pi$. Consistently, the critical parameters $m_c$ and $\theta_c$, at which the system transits back to the trivial phase for $\phi = \pi$ [Fig.~\ref{fig:3}(a,e)], satisfy the relation $m_c \approx m_0\cos{\theta_c}$, despite the presence of normal reflection and $d_{xy}$-wave AM in latter case. From the experimental perspective, we foresee that the zero-bias peak (ZBP) in scanning tunneling microscopy (STM) should be observable in generic TAJJs away from $\theta = \pi/4$ and absent in pure $d_{xy}$-wave TAJJs.

Additional topological phase diagrams as functions of chemical potential $\mu$ and SOC strength $\alpha$ are presented in SM IV \cite{Supplemental}. These results demonstrate that the chemical potential $\mu$ can be used to control the size of the topological gap, since the phase shift $E_{\text{AM}}/E_{\text{T}} \propto \sqrt{\mu}$. In summary, $d$-wave TAJJs host robust topological SC across a broad parameter range, where $d$-wave AM orientation $\theta$ serves as an internal degree of freedom that controls the topological properties by affecting the effective $d_{x^2-y^2}$-wave AM strength.

\begin{figure}[t]
    \hspace*{-0.25cm}
    \centering
    \includegraphics[scale=0.25]{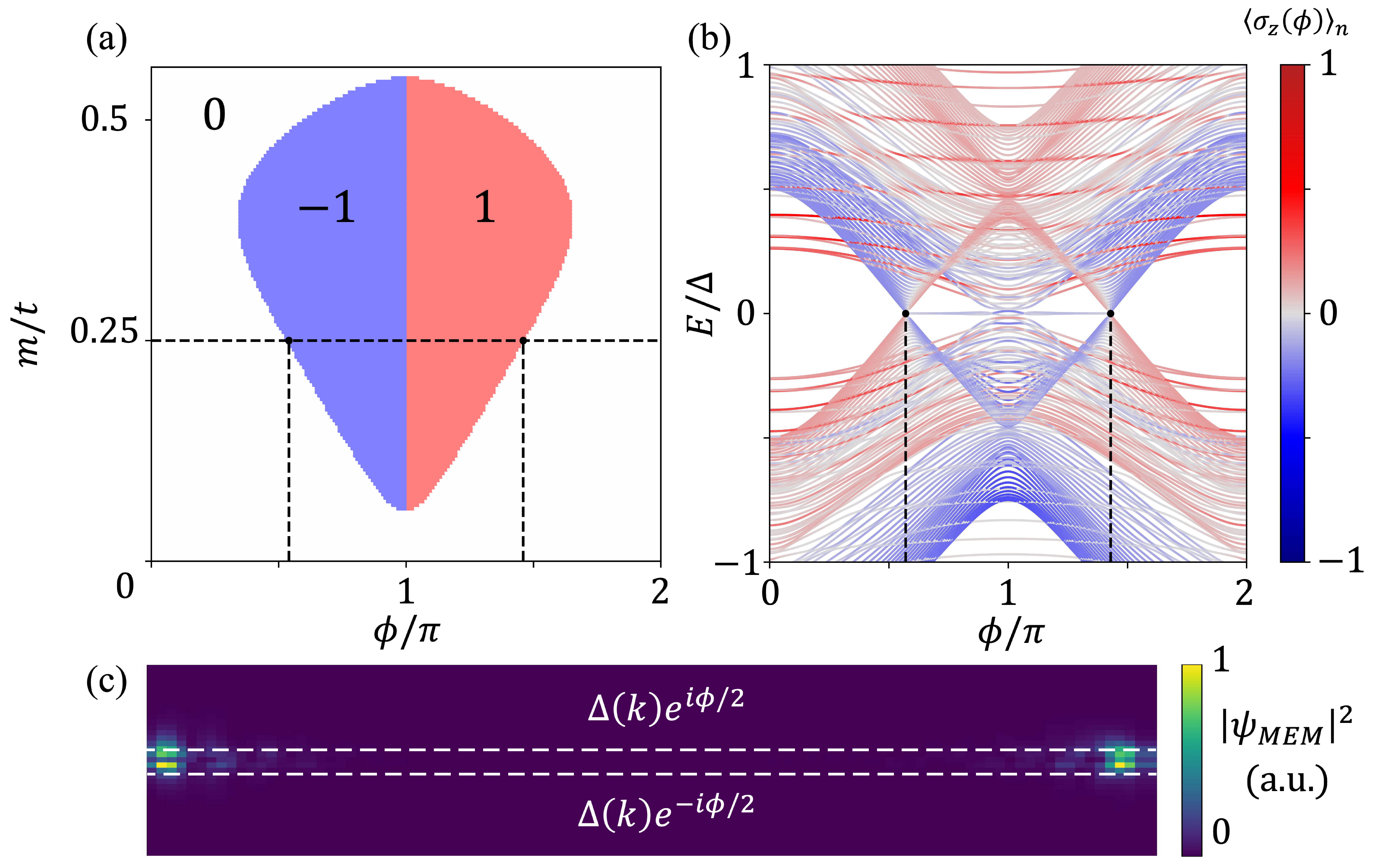}
    \caption{TAJJs in high-$T_c$ SC platform with extended $s$-wave pairing symmetry: (a) Topological phase diagram, where distinct regions are labeled by the winding number $w$.
    (b) Spin-resolved ABS spectrum along the dashed line in (a) as a function of phase difference $\phi$. 
    (c) Wavefunction distribution $|\psi_{\text{MZM}}|^2$ demonstrating MZMs localized at the junction ends (white dashed lines indicate the junction region). Here $\Delta(\bm{k})\propto (k_x^2+k_y^2)$ represents the anisotropic pairing gap and $\mu=-2.5$.}
    \label{fig:4}
\end{figure}

\emph{TAJJs in high-$T_c$ SC platforms.}\textemdash
To generalize our framework, we incorporate an anisotropic superconducting order parameter $\Delta(\bm{k})$ into TAJJs. Here we focus on $d$-wave and extended $s$-wave pairing symmetries, which are associated with the cuprate and iron-based superconductors, respectively~\cite{d_wave_gap,extended_s_gap1,extended_s_gap2}. The generalization to other pairing symmetries is straightforward.
While extended $s$-wave SC can exhibit a fully gapped quasi-particle spectrum, the nodal lines of $d$-wave SC always intersect the Fermi surface, resulting in gapless excitations \cite{Supplemental}. The interplay between $d_{x^2-y^2}$-wave AM and anisotropic SC can be described by considering the nearest-neighbor pairing Hamiltonian in SM I \cite{Supplemental}.
In this high-$T_c$ SC platform, the system $(\theta=0)$ retains BDI symmetry classification for $\Delta(\bm{k})\propto (k_x^2\pm k_y^2)$. The spin-resolved ABS spectra and topological phase diagram for the extended $s$-wave SC case are presented in Fig.~\ref{fig:4}(a, b), while the $d$-wave SC case is detailed in SM VII \cite{Supplemental}. 
For the latter, the MZMs may hybridize with bulk gapless quasi-particle excitations. However, MZMs remain distinguishable in spin-polarized STM, as these bulk zero modes are non-spin-polarized and spatially separated from MZMs \cite{Supplemental,MKP_high_Tc}. In contrast, for the fully gapped extended $s$-wave SC, MZMs emerge robustly within the topological gap, free from interference with bulk gapless excitations, and exhibit wavefunction localization at the junction ends [Fig.~\ref{fig:4}(c)]. 

Notably, other key ingredients, such as Rashba SOC and AM, can sustain high critical temperatures in experiments \cite{high_T_AM,high_T_SOC,jiang2025metallic, zhang2025crystal}, potentially enabling the realization of high-$T_c$ topological SC in TAJJs without external fields.

\emph{Conclusion and discussion.}\textemdash
We have presented a zero-field approach for realizing MZMs in TAJJs by exploiting the unique properties of $d$-wave AM. Our theoretical study reveals that the orientation angle $\theta$ of $d$-wave AM serves as an effective topological switch, as $d_{x^2-y^2}$-wave AM supports topological phases while $d_{xy}$-wave AM does not. We further extended our analysis to high-$T_c$ SC platforms by leveraging the synergy between both unconventional magnetism and SC, establishing a promising direction for designing topological zero-field devices without the detrimental effects of stray fields or external magnetic fields on SC.

While our discussion has focused primarily on $d$-wave AM for clarity, our theoretical proposal is rather general, and can be readily extended to other high even-parity wave AM (e.g., $g$ and $i$-wave AM) \cite{altermagnet_prx_1,altermagnet_prx_2}. The latest nanoscale imaging reveals the single-domain states in MnTe with characteristic scales extending up to $10 \ \mu m$ \cite{amin2024nanoscale}, which is apparently larger than the typical short junction length scale $L_{AM}\approx 100 \ nm$ \cite{topoJJ_exp_1, topoJJ_exp_2}. The spin-polarized MZMs predicted in our work could be detected using high-resolution spin-polarized STM in TAJJs and its high-$T_c$ SC variants without the significant influence from stray fields \cite{SP-STM_RMP1,SP-STM_RMP2,SP-STM_ARM,SP-STM_1,SP-STM_2,SP-STM_3}. 

Recent spin-resolved measurements have unveiled the coexistence of relativistic and non-relativistic LKSD mechanisms in MnTe \cite{LKSDexp,LKSDexp_3}, suggesting the ubiquitous nature of relativistic SOC in noncentrosymmetric altermagnets. While the 2DEG is employed here for illustrative convenience, we stress that MZMs can, in principle, be realized in TAJJs even without a 2DEG. Further investigation through experimental validation is warranted to identify optimal material candidates exhibiting the required properties \cite{LKSDexp,LKSDexp_2,LKSDexp_3,LKSDexp_4,d-wave_AM_exp,am_exp_imaging,am_ab_initio}.

\emph{Acknowledgements}\textemdash The authors thank Akito Daido for inspiring discussions. K. T. L. acknowledges the support of the Ministry of Science and Technology, China, and Hong Kong Research Grant Council through Grants No. 2020YFA0309600, No. RFS2021-6S03, No. C6025-19G, No. AoE/P-701/20, No. 16310520, No. 16307622, and No. 16309223.

\bibliographystyle{apsrev4-1}

\bibliography{References}

\newpage
\onecolumngrid 
\vspace{2em}
{\centering\Large\bfseries Supplementary Material 
for ``Topological altermagnetic Josephson junction"\par}
\vspace{1em}
{\centering Grant Z. X. Yang$^1$, Zi-Ting Sun$^{1,*}$, Ying-Ming Xie$^2$, K. T. Law$^{1,\dagger}$ \\[0.5em]
$^1$Department of Physics, Hong Kong University of Science and Technology, Clear Water Bay, Hong Kong, China \\$^2$RIKEN Center for Emergent Matter Science (CEMS), Wako, Saitama 351-0198, Japan B\par}
\vspace{1em}

\setcounter{equation}{0}
\setcounter{section}{0}
\setcounter{figure}{0}
\setcounter{table}{0}
\setcounter{page}{1}
\renewcommand{\theequation}{\arabic{equation}}
\renewcommand{\thesection}{ \Roman{section}}

\renewcommand{\thefigure}{S\arabic{figure}}
 \renewcommand{\thetable}{\arabic{table}}
 \renewcommand{\tablename}{Supplementary Table}

\renewcommand{\bibnumfmt}[1]{[S#1]}
\renewcommand{\citenumfont}[1]{S#1}

\section{I. Tight-binding model}
To obtain the spin-resolved Andreev bound-state (ABS) spectrum and topological phase diagram, we discretize the continuum Hamiltonian for topological altermagnetic Josephson junctions (TAJJs) into a tight-binding model on a square lattice as \cite{altermagnet_prx_1,altermagnet_prx_2,dwaveJJ_1,dwaveJJ_2}
\begin{equation}
    \label{H}
    H=H_0+H_{SOC}+H_{AM}+H_{SC}
\end{equation}
\begin{equation}
    H_0=-t\sum_{\langle ij, i'j'\rangle s}\left[c^\dagger_{i,j,s}c_{i',j',s'}+\text{h.c.}\right]
    -\mu\sum_{i,j,s}c^\dagger_{i,j,s}c_{i,j,s}
\end{equation}
\begin{equation}
    H_{\text{SOC}}=i\alpha\sum_{ijss'}\left[c^\dagger_{i+1,j,s}[\sigma_{y}]_{ss'}c_{i,j,s'}-c^\dagger_{i,j+1,s}[\sigma_{x}]_{ss'}c_{i,j,s'}\right] + \text{h.c.}
\end{equation}
\begin{equation}
    \begin{split}
    H_{\text{AM}}=&-m\sum_{ijss'}\left[\left(c^\dagger_{i+1,j,s}c_{i,j,s'}-c^\dagger_{i,j+1,s}c_{i,j,s'}\right)\cos{2\theta}+\left(c^\dagger_{i+1,j+1,s}c_{i,j,s'}-c^\dagger_{i+1,j-1,s}c_{i,j,s'}\right)\sin{2\theta}\right]\\
    &\times(\sigma_{z})_{ss'}\Theta(L_{\text{AM}}/2-|j|) + \text{h.c.}
\end{split}
\end{equation}
\begin{equation}
    \begin{split}
    H_{\text{SC}}=\Delta\sum_{i,j}e^{-i\text{sgn}(j)\phi/2}c_{i,j,\uparrow}c_{i,j,\downarrow}\Theta(|j|-L_{\text{AM}}/2) + \text{h.c.}
\end{split}
\end{equation}
\begin{equation}
    H^{\lambda}_{\text{SC}} = \Delta\sum_{i,j}e^{-i\text{sgn}(j)\phi/2}(c^\dagger_{i,j\pm1,\uparrow}c^\dagger_{i,j,\downarrow} + \lambda  c^\dagger_{i\pm1,j,\uparrow}c^\dagger_{i,j,\downarrow})\Theta(|j|-L_{\text{AM}}/2) + \text{h.c.},
\end{equation}
where $(i,j)$ are indices of lattice site along the $(x,y)$-direction, and $s$ represent electron spin $\{\uparrow,\downarrow\}$, 
$c_{i,j,s}$ and $c^\dagger_{i,j,s}$ are the creation and annihilation operators of an electron. $\langle\cdots,\cdots\rangle$ denotes the summation over nearest-neighbor sites, $\alpha$ is the Rashba spin-orbit coupling (SOC) strength, and Pauli matrices in spin space are $\sigma_{x,y,z}$. For the altermagnetic order parameter $H_{\text{AM}}$, $m$ is the altermagnetism (AM) strength and $\theta$ parameterizes the orientation of $d$-wave AM. $H_{\text{SC}}$ and $H_{\text{SC}}^{\lambda}$ represent isotropic $s$-wave and anisotropic superconducting order parameters, respectively. Here, $\lambda=1(-1)$ corresponds to extended $s$-wave ($d$-wave) pairing symmetries. $\Theta$ is the Heaviside step function to confine the AM and superconducting pairing terms within the junction and leads respectively, $\phi$ is the phase difference between the two superconducting leads.

The periodic boundary condition (PBC) is applied along the $x$-direction in the corresponding cases. For each $k_x$ point and fixed $\phi$, we can diagonalize the Hamiltonian $H(k_x,\phi)$ to obtain the ABS spectrum and wavefunctions.

\section{II. Winding number for the BDI symmetry class}
For the $d_{x^2-y^2}$-wave TAJJs $(\theta=0)$, the Hamiltonian (\ref{H}) satisfies the effective time-reversal, chiral, and particle-hole symmetries, leading the system to the BDI symmetry class in one-dimension, characterized by the $\mathbb{Z}$ topological invariant, the winding number $w$. The chiral symmetry operator is defined as $M_y\tau_y$ in the Nambu spinor basis $\Psi = (\psi_\uparrow, \psi_\downarrow, \psi^\dagger_\downarrow, -\psi^\dagger_\uparrow)^T$, where $M_y$ is the mirror symmetry operator defined in the main text and $\tau_y$ is the Pauli matrix in particle-hole space. Following the prescription in Ref.~\cite{winding_calculation_prl,topoJJ}, the chiral symmetry guarantees Bogoliubov-de Gennes (BdG) Hamiltonian $H(k_x)$ is block-off-diagonal in the representation of the chiral symmetry operator as
\begin{equation}
    H(k_x)=\left(
        \begin{array}{cc}
            0 & A(k_x) \\
            A^\dagger(k_x) & 0
        \end{array}
    \right)
\end{equation}
and we have $\det[H(k_x)]=\det[A(k_x)]\det[A(k_x)^\dagger]$, which implies $\det[A(k_x)]$ vanish only if $H(k_x)$ has a zero eigenvalue. Hence, the gap-closing behavior of $H(k_x)$ can be captured by the complex function 
\begin{equation}
z(k_x)=e^{i\theta(k_x)}=\frac{\det[A(k_x)]}{|\det[A(k_x)]|}
\end{equation}
The winding number $w$ is then defined as  
\begin{equation}
    w=\frac{1}{2\pi}\int_{k_x=0}^{k_x=2\pi} d\theta(k_x)=\frac{1}{2\pi i}\int_{0}^{2\pi}\frac{z'(k_x)}{z(k_x)}dk_x
    \label{winding_number}
\end{equation}
Here, the topological phase for the $d_{x^2-y^2}$-wave TAJJs is characterized by the odd winding number $w$, while the even winding number $w$ corresponds to the trivial phase.

\section{III. ABS spectrum of $d_{x^2-y^2}$-wave TAJJs: Scattering matrix formalism}
In the Nambu basis $\Psi=(\psi_\uparrow, \psi_{\downarrow},\psi_\downarrow^\dagger, -\psi_{\uparrow}^\dagger)^T$, the BdG Hamiltonian of $d_{x^2-y^2}$-wave TAJJs at $k_x=0$ can be expressed as $H_{\text{BdG}}=H_0 + H_{\text{SOC}} + H_{\text{AM}} + H_{\text{SC}}$. While the Hamiltonian in the altermagnetic weak-link region can be expressed as ($k_y\rightarrow k$): 
\begin{equation}
    \begin{split}
    & H_{\text{N}} = H_0 + H_{\text{SOC}} + H_{\text{AM}} = \left(t\hat{k}^2-\mu_s-2\alpha \hat{k}\sigma_x\right)\tau_z + m \hat{k}^2\sigma_z
\end{split}
\end{equation}
For a certain eigenenergy $E$ of the BdG equation, the corresponding momentum satisfies
\begin{equation}
    \begin{split}
        k = k_F\sqrt{1+\frac{2\alpha k \sigma_x + m k^2\sigma_z\tau_z+E\tau_z}{\mu}}
        \approx k_F\left(1+\frac{2\alpha k_F \sigma_x + m k_F^2\sigma_z\tau_z}{2\mu}\right)
        =k_F\left(1+\frac{\bm{M}(k_F)\cdot\bm{\sigma}}{2\mu}\right)
    \end{split}
    \label{momentum_relation}
\end{equation}
Since the Andreev reflection only occurs near the Fermi surface with $k_F=\sqrt{\mu_s/t}$, we neglect the energy dependence of momentum ($\mu\gg \Delta$), $\bm{\sigma}=(\sigma_x,\sigma_y,\sigma_z)$ is the Pauli matrix vector. The chemical potential $\mu_s$ is connected with the chemical potential in the tight-binding model by the relation $\mu_s = 4t + \mu$. To illustrate the spin-momentum locking due to both AM and Rashba SOC, the spin-polarization vector is defined as
\begin{equation}
    \bm{M}(k) = \left(2\alpha k, 0, m k^2\tau_z\right)
\end{equation}
Here we note that the in-plane and out-of-plane spin polarizations for the electron and hole states are same and opposite, respectively, as $\left[M_e(k)\right]_{x/y}=\left[M_h(k)\right]_{x/y}$ and $\left[M_e(k)\right]_{z}=-\left[M_h(k)\right]_{z}$. Therefore, the Rashba SOC terms solely do not contribute to the momentum transfer during the Andreev reflection at $k_x=0$ (in the first order of Eq. \ref{momentum_relation}), only the AM term contributes to the momentum transfer as
\begin{equation}
    k_{e\rightarrow h} = k_e - k_h = \frac{m}{t}k_F\sigma_z.
\end{equation}
Consequently, although spin is not a good quantum number because of Rashba SOC, we can still use out-of-plane spin polarization $\nu=\pm 1$ to label the spin-up/down momentum $k_{e,h}^{\nu}$. To obtain the analytical expression for the ABS spectrum, here we adopt the Andreev approximation and neglect the normal reflection. The $S$-matrix is defined as $\Psi^\nu_{\text{out}}=S\Psi^\nu_{\text{in}}$, where the incoming wave and outgoing wave have the form $\Psi^\nu_{\text{in}}=\left(a^{\nu,+}_e(L),a^{\nu,-}_h(L),a^{\nu,-}_e(R),a^{\nu,+}_h(R)\right)^T$ and $\Psi^\nu_{\text{out}}=\left(a^{\nu,-}_e(L),a^{\nu,+}_h(L),a^{\nu,+}_e(R),a^{\nu,-}_h(R)\right)^T$. By matching the boundary conditions of the wavefunction and its derivative, we can have 
\begin{equation}
    S=\left(
        \begin{array}{cc}
            S_A(-\phi/2)&\\
            &S_A(\phi/2)\\
        \end{array}
    \right), \
    S_A(\phi)= e^{i\alpha}\left(\begin{array}{cc}
        0 & e^{i\phi}\\
        e^{-i\phi} & 0
    \end{array}\right)
\end{equation}
Here $\alpha=\arccos{(E/\Delta)}$. Meanwhile, we can obtain the $T$-matrix from $\Psi^\nu_{\text{in}}=T \Psi^\nu_{\text{out}}$ and definitions of coefficient.  
\begin{equation}
    T = \left(\begin{array}{cccc}
        0&0&e^{-ik_e^\nu L_{\text{AM}}}&0\\
        0&0&0&e^{ik_h^\nu L_{\text{AM}}}\\
        e^{-ik_e^\nu L_{\text{AM}}}&0&0&0\\
        0&e^{ik_h^\nu L_{\text{AM}}}&0&0
    \end{array}\right)
\end{equation}
The ABS bound state spectrum is given by the equation $\det\left[1-ST\right]=0$, which yields
\begin{equation}
    E = \Delta \cos\left(\frac{\phi}{2}- \frac{m}{2t}k_F L_{\text{AM}}\sigma_z\right)=\Delta \cos\left(\frac{\phi}{2}-\frac{\pi}{2}\frac{E_{AM}}{E_{\text{T}}}\sigma_z\right)
\end{equation}
where the phase shift $E_{\text{AM}}= m k_F^2$ induced by the AM is given in units of Thouless energy $E_T=(\pi/2)v_F/L_{\text{AM}}$. At the same time, the system possesses chiral symmetry defined as $C = M_y \tau_y$ as discussed in the main text. Under the chiral symmetry operation, the out-of-plane spin and Hamiltonian follow the relation $C \sigma_z C^{-1}=-\sigma_z$ and $C H(\phi) C^{-1} = -H(\phi)$, which leads to the relation of ABS spectrum $E_{\nu}(\phi)=-E_{-\nu}(\phi)$. The analytical ABS spectrum for the $d_{x^2-y^2}$-wave TAJJs at $k_x=0$ is obtained readily now
\begin{equation}
    E_{\nu}(\phi) = \pm\nu \Delta \cos\left(\frac{\phi}{2}\mp\frac{\pi}{2}\frac{E_{AM}}{E_T}\nu\right)
\end{equation}
The zero-energy solution is given by 
\begin{equation}
    \frac{\phi}{2} - \frac{\pi}{2}\frac{E_{AM}}{E_T}\nu = \frac{\pi}{2} + n\pi, \ n \in \text{integer}
\end{equation}
However, the reasonable strength of the AM experimentally should be smaller than the kinetic energy term as $t > m$.

\section{IV. Evolution of topological phases with chemical potential and SOC strength}
For the TAJJs in BDI symmetry class $(\theta=0)$, we calculate the topological phase diagrams characterized by the winding number $w$ in the parameter space $(\phi,\mu)$ and $(\phi,\alpha)$, as shown in Fig. \ref{additional_phase_diagram} (b) and (c). The parameters used here are identical to those in the main text.

In Fig. \ref{additional_phase_diagram} (b), system remains topological around $\phi=\pi$ across nearly the entire chemical potential range $\mu\in[-3,3]$. As $\mu$ approaches half-filling, additional sub-regions emerge and the topological region expands. The phase diagram exhibits particle-hole symmetry, manifested by the symmetry with respect to $\mu=0$. The corresponding ABS spectrum, presented in Fig. \ref{additional_phase_diagram} (a), features Majorana zero modes (MZMs), highlighted by red lines. 

Our proposed TAJJ framework requires a finite Rashba SOC to stabilize the topological phase. In the absence of Rashba SOC $(\alpha=0)$, the system remains topologically trivial, since the system needs the Rashba SOC effect to protect the topological gap at $k_x\ne 0$. The non-zero Rashba SOC $(\alpha\ne0)$ drives the system into a topological phase. Notably, the topological phase is remarkably robust against variations in the SOC strength $\alpha$. This robustness suggests a simplification on experimental realization, we can utilize the intrinsic SOC of altermagnet, without the need for fine-tuned external SOC engineering. 
\begin{figure}[h]
    \centering
    \includegraphics[scale=0.35]{./pic/additional_phase_diagram.jpg}
    \caption{(a) ABS spectrum as a function of chemical potential $\mu$ at a fixed phase difference $\phi=0.6\pi$. The red lines highlight the MZMs. (b) and (c) correspond to the topological phase diagram in the parameter space $(\phi, \mu)$ and $(\phi,\alpha)$, respectively. The winding number $w$ is calculated using (\ref{winding_number}). The regions with white color represent the topologically trivial phase.}
    \label{additional_phase_diagram}
\end{figure}

\section{V. The ABS spectrum degeneracy and vanishing spin-polarization in $d_{xy}$-wave TAJJs}
\begin{figure}[h]
    \centering
    \includegraphics[scale=0.4]{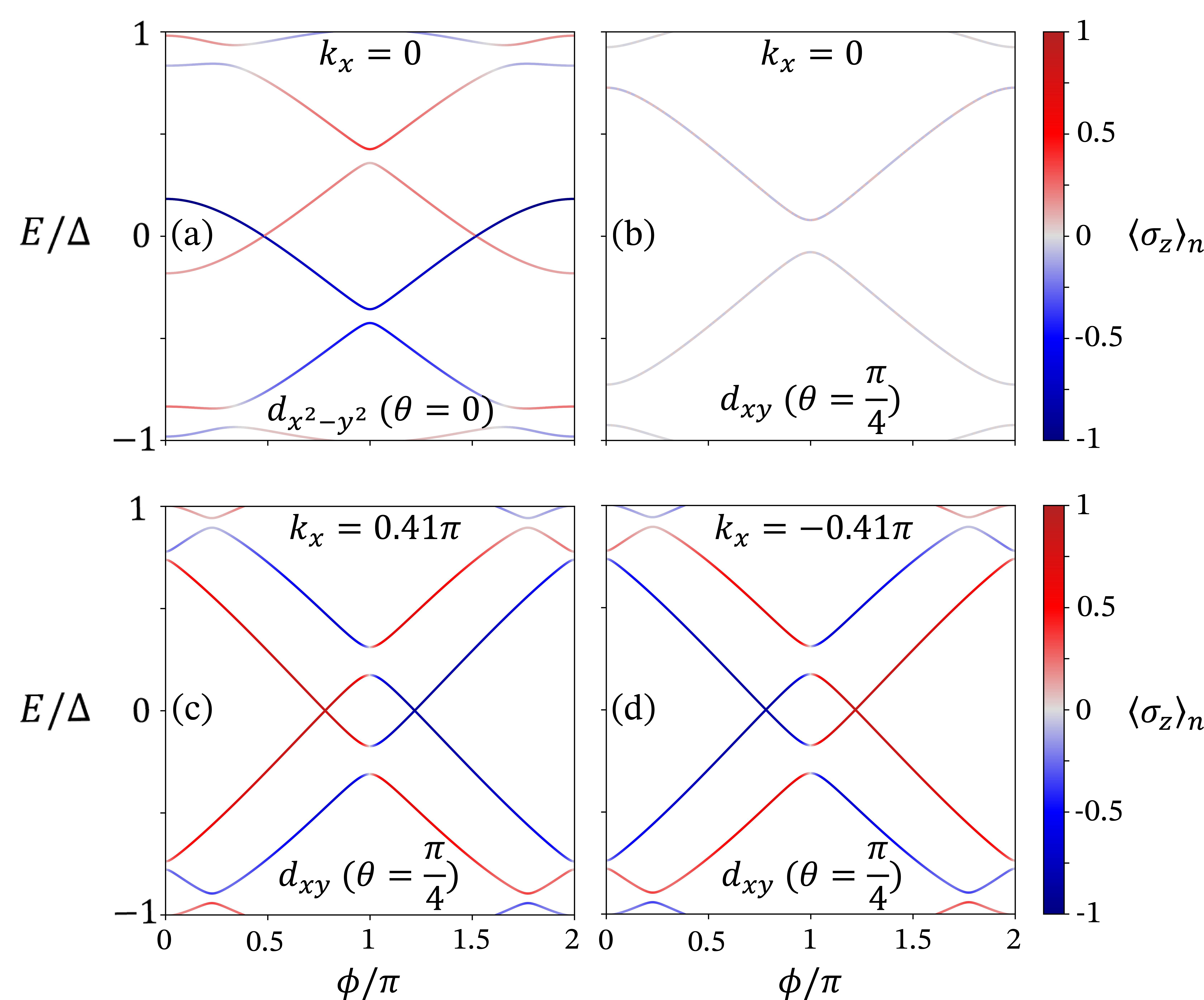}
    \caption{ABS spectra of planar TAJJs for selected $k_x$ modes under PBC. The color scale indicates the normalized spin-polarization of each ABS. The emergence of topology for TAJJs is governed by band inversion at $k_x=0$. (a) The ABS spectrum of $d_{x^2-y^2}$-wave TAJJs at $k_x=0$: A topological gap opens as the phase difference $\phi$  around $\pi$. (b-d) ABS spectra of $d_{xy}$-wave TAJJs for distinct $k_x$ modes. The altermagnetic order parameter vanishes $M(\theta=\pi/4)$ at $k_x=0$ and the Kramers spin degeneracy is preserved. (c, d) Opposite spin-polarizations emerge at $\pm |k_x|$, resulting in net spin cancellation for the $d_{xy}$-wave TAJJs.}
    \label{ABS_pbc}
\end{figure}

The order parameter for the general $d$-wave altermagnetism is given by:
\begin{equation}
    M(\theta)\sigma_z = m\left[(k_x^2-\hat{k}_y^2)\cos{2\theta}+4k_x\hat{k}_y\sin{2\theta}\right]\sigma_z. 
\end{equation}
For $d_{x^2-y^2}$-wave TAJJs $(\theta=0)$, the system can exhibit a topological phase characterized by an odd winding number $w$ when $\phi$ is around $\pi$. In contrast, $d_{xy}$-wave TAJJs $(\theta=\pi/4)$ are topologically trivial due to a vanishing altermagnetic order parameter at $k_x=0$, which enforces Kramers spin-degeneracy of ABSs [Fig. \ref{ABS_pbc}(a), (b)]. 

The $d_{xy}$-wave case further satisfies $M(\theta=\pi/4)\sigma_z = 4m k_x\hat{k}_y\sigma_z$, implying degeneracy between ABSs $\ket{\psi_n(k_x,\sigma_z)}$ with opposite out-of-plane spin-polarization at $\pm |k_x|$, as illustrated in Fig. \ref{ABS_pbc}(c), (d). In a planar Josephson junction, the total ABS at a given energy level is a superposition of these degenerate states:
\begin{equation}
    \ket{\psi_n(\theta=\frac{\pi}{4},\phi)}=\frac{1}{\sqrt{2}}\left(\ket{\psi_n(k_x,\uparrow,\phi)}+\ket{\psi_n(-k_x,\downarrow,\phi)}\right).
\end{equation}
where the supercurrent flows along the nodal lines. Despite the spin-polarization for individual modes with $k_x\ne0$ remain finite, the net spin-polarization $\langle\sigma_z(\phi)\rangle_n=\bra{\psi_n(\phi)}\sigma_z\ket{\psi_n(\phi)}$ vanishes due to this superposition for the $d_{xy}$-wave TAJJs. This conclusion can be generalized to other cases with distinct even-parity altermagnetism. 

For the $d_{x^2-y^2}$-wave TAJJs, degeneracy at $k_x\ne0$ occurs between states with identical spin polarization instead, as $\ket{\psi_n(k_x,\uparrow(\downarrow))}$ and $\ket{\psi_n(-k_x,\uparrow(\downarrow))}$, since $M(\theta=0)\sigma_z = m (k^2_x-\hat{k}^2_y)\sigma_z$. Thereby, superposition here enhances the net spin-polarization. 
In generic cases ($\theta\ne 0, \pi/4$), all such degeneracies are lifted. 

\section{VI. Quasiparticle excitations and trivial zero-modes in nodal superconductors}
In the superconducting leads with the presence of a strong Rashba SOC effect, the quasiparticle excitations in the momentum space are given by 
\begin{equation}
    E_\pm(\bm{k}) = \sqrt{\epsilon_{\pm}^2(\bm{k}) + \Delta^2(\bm{k})}
\end{equation}
where the normal state energy is $\epsilon_\pm(\bm{k})=-2t(\cos{k_x}+\cos{k_y})-\mu \pm 2\alpha \sqrt{\sin^2{k_x}+\sin^2{k_y}}$. The anisotropic pairing potential is denoted as $\Delta(\bm{k})$, where the extended $s$-wave and $d$-wave pairing potentials are given by $\Delta(\cos{k_x}+\cos{k_y})$ and $\Delta(\cos{k_x}-\cos{k_y})$ respectively. The Fermi surface and nodal lines are determined by the conditions $\epsilon_{\pm}(\bm{k})=0$ and $\Delta(\bm{k})=0$. Consequently, the crossing points $\bm{k}_{0}$ between the Fermi surface and nodal lines indicate the presence of zero-energy quasiparticle excitations $E_{\pm}(\bm{k}_0)=0$, which is not protected by the topology, as illustrated in Fig. \ref{nodalSC_FS}. For the $d$-wave pairing symmetry, the zero-energy states always exist no matter the size of the Fermi surface, and may hybridize with the MZMs in the TAJJs as the trade-off for a high critical temperature. However, these trivial zero-energy states are non-spin-polarized and spatially separated from MZMs. More importantly, for the extended $s$-wave pairing symmetry, there are no trivial zero-energy states, as long as the Fermi surface stays away from the nodal lines.
\begin{figure}[h]
    \centering
    \includegraphics[scale=0.37]{./pic/nodalSC_FS.jpg}
    \caption{The Fermi surface contours (red and blue curves) of superconducting leads with proximitized Rashba SOC at different chemical potential and nodal lines (dashed lines) of anisotropic pairing potential $\Delta(\bm{k})$. (a-b) and (c-d) correspond to extended $s$-wave and $d$-wave superconductivity respectively. The nodal lines are determined by the condition $\Delta(\bm{k})=0$. The crossing points between the Fermi surface and nodal lines indicate the presence of trivial zero-energy quasiparticle excitations.}
    \label{nodalSC_FS}
\end{figure}

\section{VII. TAJJs in $d$-wave superconductivity platforms}
In the main text, we have already discussed the realization of topological superconductivity (SC) in TAJJs with a higher critical temperature, by incorporating SC with anisotropic pairing symmetries (extended $s$-wave and $d$-wave). Here we provide a more detailed discussion on the $d$-wave SC platforms.
\begin{figure}[h]
    \centering
    \includegraphics[scale=0.3]{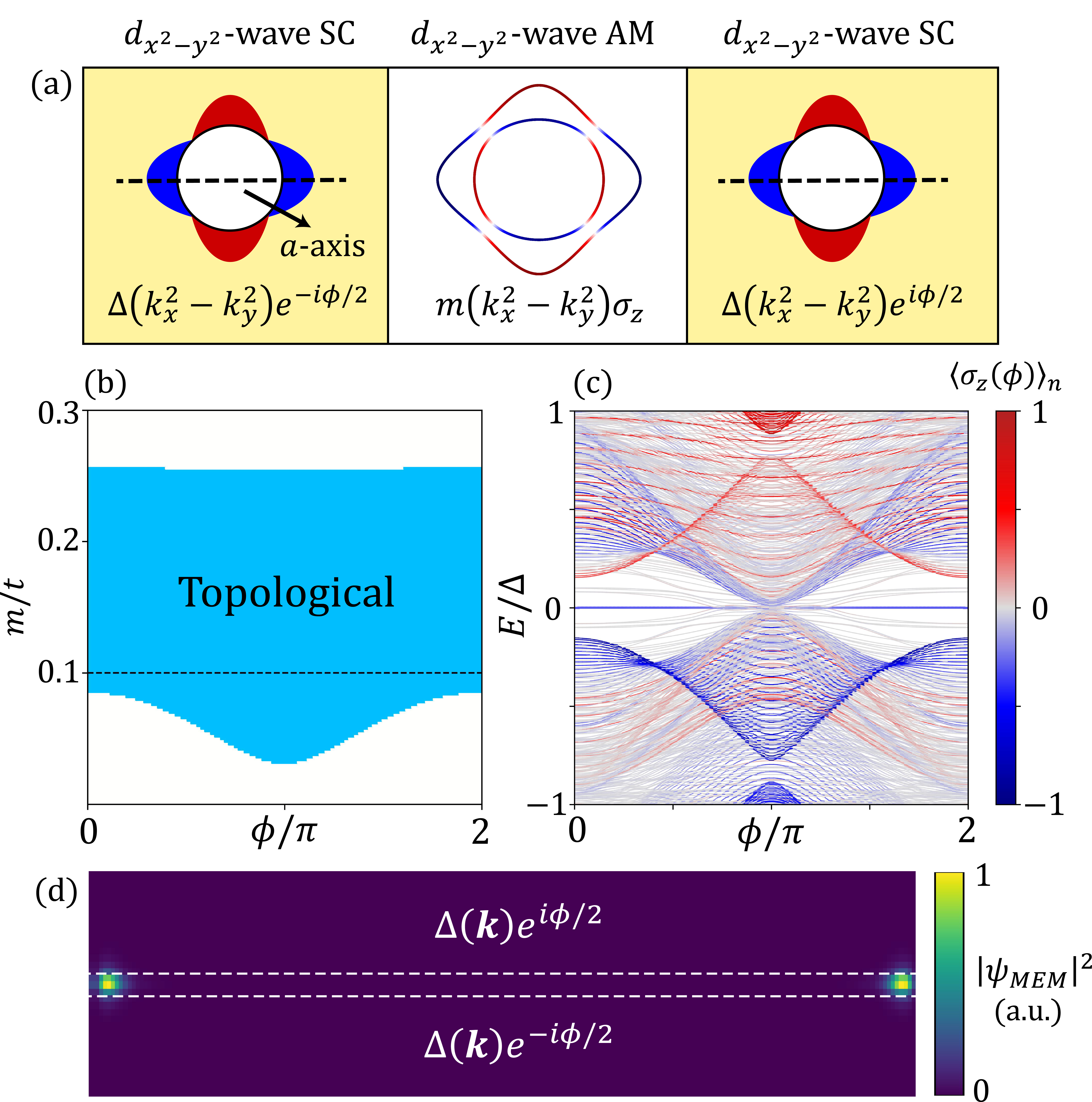}
    \caption{ (a) Schematic of TAJJs implemented in high-$T_c$ SC platforms with $d_{x^2-y^2}$-wave SC leads. (b) Topological phase diagram, where regions with odd winding numbers indicate topological phases. (c) Spin-polarized ABS spectrum along the dashed line in (b) as a function of phase difference $\phi$, with
 spin-up (spin-down) states shown in red (blue). The nearly zero in-gap states originating from the nodes of $d$-wave SC remain non-spin-polarized (in gray). (d) Probability density distribution $|\psi_{\text{MZM}}|^2$ demonstrating MZMs localized at the junction ends (white dashed lines indicate the junction region). $\Delta(\bm{k})$ represents the pairing gap with $d$-wave pairing symmetry.}
    \label{d-wave_sc}
\end{figure}

To ensure a non-vanishing gap along the supercurrent direction, we arrange the system so that the $a$-axes of the two $d_{x^2-y^2}$-wave superconductors are aligned along the current direction \cite{dwaveJJ_1}, as demonstrated in Fig.~\ref{d-wave_sc} (a). In such a setup, the system maintains its BDI symmetry classification.  Remarkably, as shown in the topological phase diagram in Fig. \ref{d-wave_sc} (b), MZMs emerge independently of the phase difference $\phi$ over a much broader range of AM strength $m$ compared to the $s$-wave SC case. The dashed line in the phase diagram corresponds to the spin-polarized ABS spectrum shown in Fig. \ref{d-wave_sc} (c), where spin-polarized MZMs appear at zero energy without requiring fine-tuning of $\phi$.

Unlike $s$-wave SCs, the nodal lines of $d$-wave pairing potential always have intersections with the Fermi surface as discussed in the last section, rendering additional nearly-zero modes in the ABS spectrum. However, these modes remain non-spin-polarized (represented by gray color in Fig.~\ref{d-wave_sc}(c)), in stark contrast to the spin-polarized MZMs. On the other hand, the MZMs themselves are localized at the ends of the junction, as demonstrated by the wavefunction distribution $|\psi_{\text{MZM}}|^2$ in Fig.~\ref{d-wave_sc}(d). For $d$-wave superconductor platforms, the spin-polarized nature of the MZMs provides a clear signature to distinguish them from the non-spin-polarized nearly-zero modes arising from the $d$-wave nodes.

\section{VIII. Edge magnetization in the TAJJs}
Altermagnets feature the zero net magnetization in the bulk, akin to the antiferromagnet. However, under open boundary conditions (OBC), edge magnetization is permitted. To characterize this behavior, we define the local magnetization profile, which quantifies the out-of-plane magnetization at each lattice site as 
\begin{equation}
    \langle\sigma_z^i(\phi)\rangle = \sum_{E_n<0}\bra{\psi_n(\phi)}\sigma_z^i\ket{\psi_n(\phi)}
\end{equation}
Here, $\sigma_z^i$ represents the out-of-plane spin operator at site $i$, and $\ket{\psi_n(\phi)}$ denotes the eigenstate of BdG Hamiltonian. The summation is performed over all eigenstates with negative energies. 

\begin{figure}[h]
    \centering
    \includegraphics[scale=0.4]{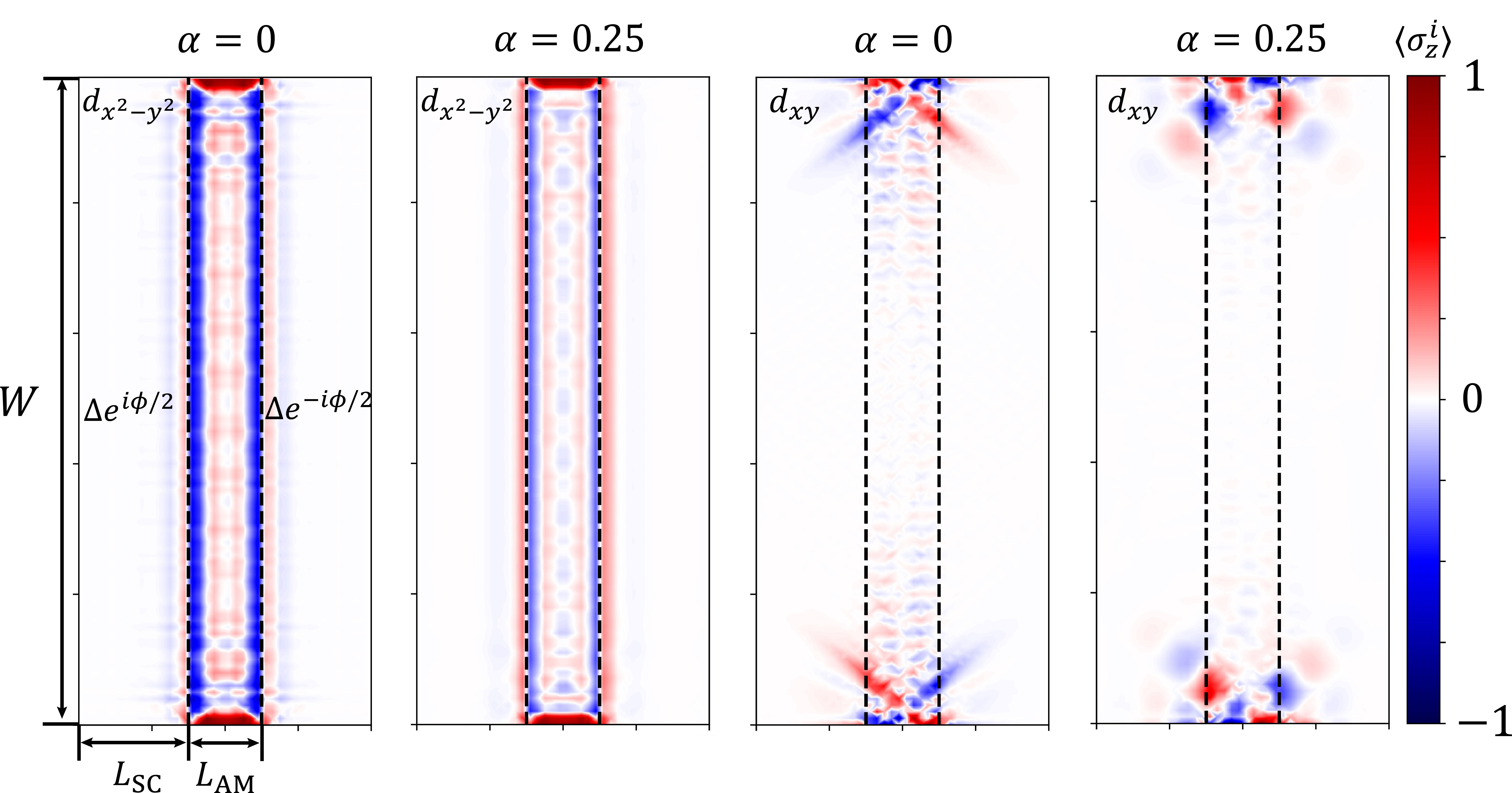}
    \caption{Local magnetization profile of TAJJs in the presence $(\alpha\ne0)$ and absence $(\alpha=0)$ of Rashba SOC, respectively. The parameters used are $(m=0.1,\phi=0.6\pi, W=100, L_{\text{AM}}=5, L_{\text{SC}}=15)$. The dashed line regions demarcate the altermagnetic weak-link regions. The color bar represents the normalized out-of-plane spin polarization, illustrating the spatial magnetization distributions across the junctions.}
    \label{local_spin}
\end{figure}

As illustrated in Fig. \ref{local_spin}, the out-of-plane magnetization $\langle\sigma_z^i\rangle$ exhibits distinct spatial distributions. For the $d_{x^2-y^2}$-wave TAJJs, the magnetization is predominantly distributed along the boundaries of the junctions. In contrast, for the $d_{xy}$-wave TAJJs, the magnetization is localized at the corners of the junctions. As we elucidated before, the degenerate states possess opposite out-of-plane spin polarizations, rendering the cancellation of spin polarization in the ABSs for the $d_{xy}$-wave TAJJs. Consequently, the bulk regions of the $d_{xy}$-wave TAJJs are entirely non-magnetic. On the other hand, the spin-polarized eigenstates in the $d_{x^2-y^2}$-wave TAJJs lead to a small but finite magnetization in the bulk. 

Furthermore, we sum over the local magnetization profiles over all lattice sites, defining the total out-of-plane magnetization as $\langle\sigma_z\rangle=\sum_i \langle\sigma_z^i\rangle$. This quantity remains finite for the $d_{x^2-y^2}$-wave TAJJs but vanishes for the $d_{xy}$-wave TAJJs. In summary, we conclude that the bulk out-of-plane magnetization vanishes when a supercurrent flows along the nodal line of altermagnet. This current-tuned magnetization in the bulk, combined with edge magnetization, underscores the anisotropic transport properties of TAJJs with different $d$-wave AM orientations.   

\section{IX. Asymmetry of ABS spectrum in $d_{xy}$-wave TAJJs under OBC}
\begin{figure}[h]
    \centering
    \includegraphics[scale=0.3]{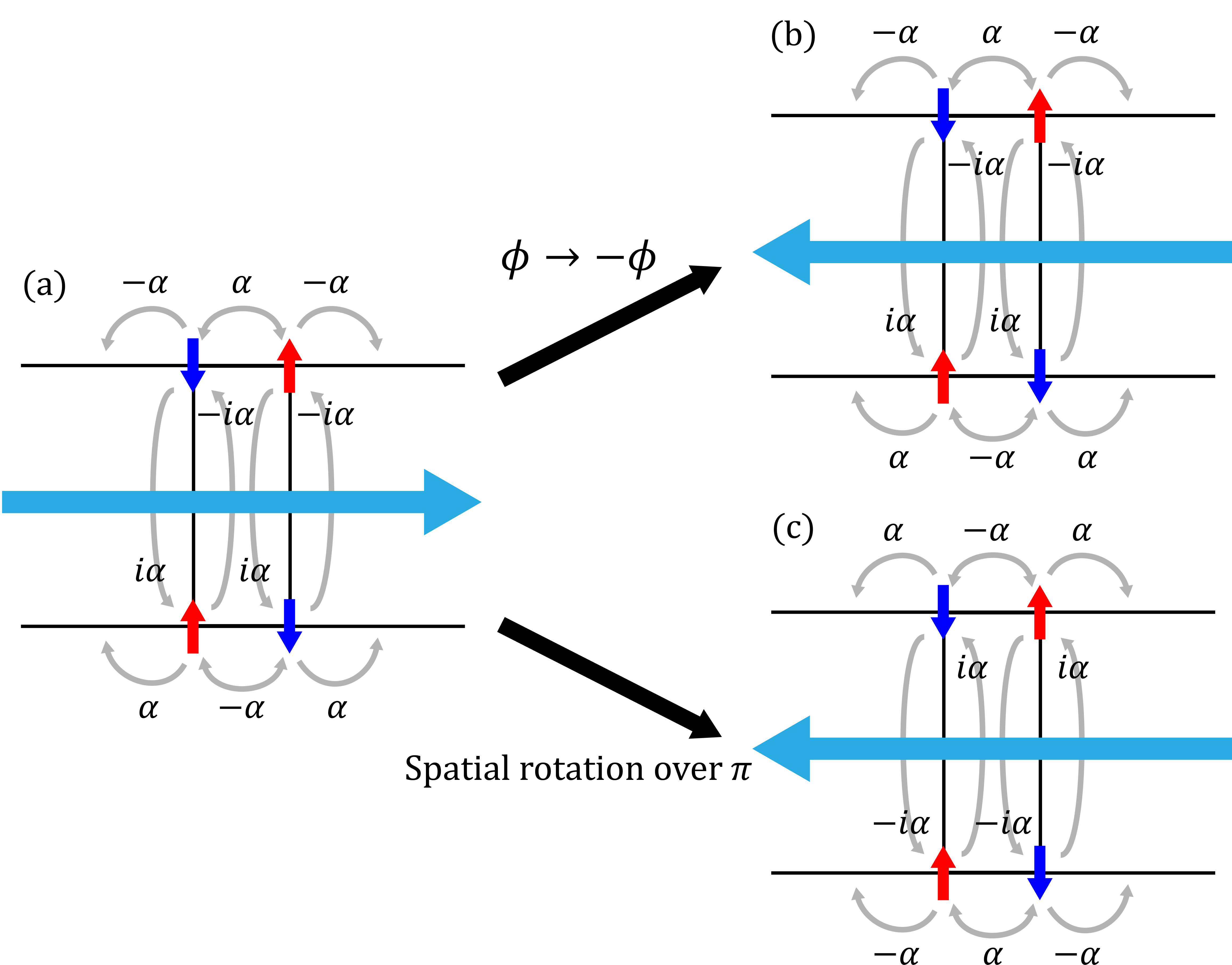}
    \caption{Schematic illustration of edge diode effect in the $d_{xy}$-wave TAJJs. The light blue arrows indicate the supercurrent direction, and the blue and red arrows represent out-of-plane edge magnetization at junction corners. The hopping of Rashba SOC terms is denoted. Fig. (c) corresponds to Fig. (a) after an in-plane rotation over $\pi$ spatially, while Fig. (b) hosts a reverse supercurrent compared with Fig. (a). The inequivalence between Fig. (b) and (c) reveals the edge diode effect in the $d_{xy}$-wave TAJJs under OBC.}
    \label{edge_diode}
\end{figure}

In the main text, we demonstrate that the ABS spectrum of $d_{xy}$-wave TAJJs under OBC exhibits an asymmetry near zero energy with respect to the phase difference $\phi$. Here we provide an explanation of this phenomenology. This asymmetry can lead to edge diode effect, as a direct consequence of non-zero edge magnetization localized at the $d_{xy}$-wave altermagnetic junction corners [Fig. (\ref{local_spin}), (\ref{edge_diode})]. The underlying mechanism can be understood as follows: The Rashba SOC terms, $\sigma_{x/y}$, are off-block-diagonal in spin space, which only mediate hopping between electrons with opposite spins. Therefore, combined with corner-localized edge magnetization in $d_{xy}$-wave junctions, the inversion and time reversal symmetries are broken simultaneously on the edges. This enables the inequivalence between a junction experiencing a spatial rotation over $\pi$ and one with reverse supercurrent [Fig. \ref{edge_diode}].

By contrast, under PBC, the bulk altermagnet exhibits zero net magnetization, suppressing the edge magnetization entirely. Furthermore, in $d_{x^2-y^2}$-wave junctions, edge magnetization is distributed with uniform out-of-plane spin along the boundaries, preventing Rashba SOC-mediated asymmetry and thus no inversion symmetry breaking on the boundaries. This analysis underscores that the ABS asymmetry and its associated edge diode effect are unique to $d_{xy}$-wave TAJJs under OBC.

\end{document}